\documentclass[conf]{new-aiaa}
\usepackage[utf8]{inputenc}
\usepackage{algorithm}
\usepackage{algpseudocode}
\usepackage{graphicx}
\usepackage{subcaption}
\usepackage{amsmath}
\usepackage[version=4]{mhchem}
\usepackage{siunitx}
\usepackage{longtable,tabularx}
\usepackage{booktabs}
\title{GPU-Accelerated Simulations of Moving Boundary Problems and Fluid-Structure Interaction at Extreme Scales}

\author{Sushrut Kumar\footnote{Graduate Student.}, Jung-Hee Seo\footnote{Associate Research Professor.} and Rajat Mittal.\footnote{Professor, AIAA Associate Fellow.}}
\affil{Department of Mechanical Engineering, Johns Hopkins University, MD, USA}
\author{Joshua Romero\footnote{Senior Software Engineer} and Massimiliano Fatica.\footnote{Senior Director}}
\affil{NVIDIA Corporation, CA, USA}

\begin{document}

\maketitle

\begin{abstract}
Computational fluid dynamics and fluid-structure interaction simulations involving moving and deforming bodies is extremely hard. In this work, we present a graphical processing unit (GPU) optimized implementation of the sharp-interface immersed boundary method. The method allows performing simulation around complex stationary as well as moving bodies on a Cartesian grid. We base our implementation on the ViCar3D framework and make use of OpenACC, CUDA, NCCL and MPI. We test the implementation across grid sizes ranging from $O(10 \,\text{million})$ to $O(1 \, \text{billion})$ points and achieved a 20X speedup compared to existing CPU implementation. We next present our multi-GPU implementation by utilizing CUDA streams and NCCL communicators. This enables us to obtain a >90\% strong and weak scaling efficiencies. Next we demonstrate the capability of the developed software to simulate a turbulent fluid flow and coupled fluid-structure interaction in flapping bat wing in flight at Re=5000. 
\end{abstract}



\section{Introduction}
\lettrine{C}{omputational} fluid dynamics (CFD) has become a cornerstone of modern engineering and scientific discovery, spanning fields from aerospace design and renewable energy to complex biological systems and medical diagnostics. As we push the boundaries of these disciplines, the systems of interest increasingly involve multi-scale physics and highly complex, moving geometries. In this context, the ability to accelerate simulations is not merely a matter of convenience; it is a prerequisite for achieving the high-fidelity resolutions such as Large Eddy Simulation (LES) and Direct Numerical Simulation (DNS). It is necessary to uncover fundamental physics and drive iterative design cycles. To meet this demand, numerical algorithms must be re-engineered to fully exploit the massive parallelism offered by modern high-performance computing (HPC) architectures.

Historically, the CFD landscape has been dominated by body-fitted mesh methods. In these approaches, the computational grid is meticulously deformed to conform to the surface of the immersed body. While highly accurate for static or small-displacement problems, the generation of such meshes becomes a significant bottleneck when dealing with the extreme deformations seen in soft-tissue biological systems or complex flapping flight \cite{mittal2023origin, viola2022fsei}. For instance, the simulation of a bat wing involves not just movement, but a radical topological changes such as folding and stretching of wing. Maintaining grid quality in a body-fitted framework under these conditions often requires frequent, computationally expensive remeshing, which can introduce numerical diffusion and severely limit the stability of the solver.

Immersed boundary methods (IBM) have emerged as a robust alternative by decoupling the fluid grid from the solid geometry. By simulating complex bodies on a fixed Cartesian grid, IBMs bypass the meshing bottleneck entirely. These methods are generally bifurcated into diffuse-interface and sharp-interface approaches. While diffuse-interface methods initially pioneered by Peskin \cite{Peskin1977-heart} are widely used for their simplicity, they rely on regularized delta functions that spread the boundary influence over multiple grid cells. This "smearing" effect can compromise the accuracy of near-wall flow physics, particularly in high-Reynolds-number regimes where boundary layer resolution is critical \cite{mittal2023origin}. Conversely, sharp-interface IBMs, such as the ghost-cell method, enforce no-slip and no-penetration conditions directly at the boundary. This provides the accuracy and flexibility required for the high-fidelity aeroelastic simulations of flapping-wing systems \cite{mittal2025freeman}.

The need for such high-fidelity tools is perhaps most evident in the study of natural fliers. Over millions of years, evolution has refined the flight of birds, insects, and bats into a masterclass of agile locomotion. Among these, bats are particularly unique; they are the only mammals capable of powered flight, utilizing a highly articulated skeletal hand-wing draped in a thin, elastic membrane. Unlike the stiff wings of an aircraft or even the relatively constrained wings of birds, a bat's wing undergoes massive geometric deformation through a complex interplay of pitching, flapping, and active skeletal articulation \citep{norberg1987ecological, swartz2015advances}. This primary challenge is to model highly flexible and zero-thickness membrane interacting with a vortex-dominated flow field while assimilating data from experiments.

Understanding the coupled aero-structural dynamics of bat flight has historically been limited by experimental constraints. While early studies provided invaluable data on physiology and kinematics \citep{riskin2008quantifying}, and later Particle Image Velocimetry (PIV) work illuminated wake structures \citep{muijres2008leading}, experimentalists often struggle to capture flow data in the immediate vicinity of a rapidly deforming wing. Computational modeling offers a path forward, yet early numerical efforts often relied on one-way coupling or simplified "bat-inspired" kinematics that ignored the full articulation of the finger joints \citep{viswanath2014straight}. Recent FSI models have begun to incorporate hyperelastic materials and skeletal bones \citep{li2019novel, lauber2023rapid}, but many still omit critical segments like the propatagium or the specific joint articulations that define the bat’s unique maneuverability.

The computational cost of resolving such a detailed, fully coupled FSI problem is immense. Traditionally, CFD codes were built for Central Processing Units (CPUs) using Message Passing Interface (MPI) for distributed scaling. However, the rise of Graphics Processing Units (GPUs), originally driven by the gaming and AI industries, has shifted the paradigm of scientific computing. GPUs offer a data parallel architecture that is perfectly suited for the structured Cartesian grids used in IBM. Yet, porting legacy codes to a GPU-native environment is fraught with challenges, particularly regarding memory bound operations and the divergent logic required for sharp-interface boundary treatments. Despite these hurdles, the potential for order-of-magnitude speedups makes GPU acceleration an essential frontier for simulating complex biological fliers \cite{raj2023gpu, viola2023gpu}.

A significant gap currently exists in the literature: there is a lack of fully GPU-native, scalable solvers capable of handling the extreme FSI required to simulate a bat wing with biological realism. Most existing GPU solvers are either tailored for rigid bodies or lack the multi-regime flexibility needed for broad application. This work aims to bridge that gap. We present a GPU-accelerated sharp-interface immersed boundary solver designed specifically for high-fidelity FSI simulations. 

By applying this solver to the case of a bat wing in forward flight, we attempt to recapitulate the complex aero-structural dynamics of the \emph{Pteropus pumilus} with unprecedented realism. Our model includes the full wing anatomy including both, the arm armwing and handwing segments. Here, the joing motion is obtained from high-speed videogrammetry and assimilated within simulation to improve realism. We investigate the specific mechanisms within the flow and membrane dynamics that dictate aerodynamic performance, seeking to understand the role that fluid-structure interaction play in enabling agile flight. Ultimately, this work provides both a computational framework for next-generation GPU-CFD and a look into the mechanics that allow the bat to fly efficiently.

\section{Computational Methods}

\subsection{Governing Equations and Numerical Methods}
The numerical methodology is centered on the concurrent solution of fluid flow and structural deformation. We perform Direct Numerical Simulations (DNS) of incompressible flow by solving the conservative form of the 3D, time-dependent Navier-Stokes equations, as shown in Equation \ref{eq:NS}:

\begin{equation}
    \frac{\partial u_i}{\partial t} + \frac{\partial u_i u_j}{\partial x_j} = - \frac{\partial p}{\partial x_j} +  \frac{1}{\text{Re}}\frac{\partial^2 u_i}{\partial x_j^2}; \, \, \, \frac{\partial u_i}{\partial x_i}=0 
    \label{eq:NS}
\end{equation}

The solver utilizes a fractional step method \cite{CHORIN196712} to decouple the velocity and pressure fields. This involves splitting Equation \ref{eq:NS} into an advection-diffusion equation and a pressure Poisson equation. For spatial discretization, we employ second-order finite-difference schemes on a non-conformal Cartesian grid. To solve the resulting systems, we implement a scheduled relaxed Jacobi (SRJ) method for the advection-diffusion components and a BiCGStab solver, preconditioned with SRJ, for the pressure Poisson equation. 

To handle the complex, moving boundaries of the wing, we employ a sharp-interface immersed boundary method. While the physical Reynolds number for a bat in flight is approximately $Re \approx 6 \times 10^4$, resolving such high values for deforming surfaces is computationally prohibitive. Following established literature on flapping wings, we maintain a simulation Reynolds number of $1,000$ while matching the Strouhal number ($St = 0.25$), ensuring that the primary vortex dynamics and boundary layer physics remain representative of the full-scale flight regime.

\subsection{Bat Wing Geometry and Kinematics}
The wing model is based on the biological morphology of the \emph{Pteropus pumilus}. Unlike rigid or simple flapping wings, the bat wing is highly articulated, comprising four primary segments supported by a skeletal structure: the propatagium (W1), plagiopatagium (W2), dactylopatagium major (W3), and dactylopatagium medius (W4). 

To ensure realistic flapping, wing kinematics are derived from high-resolution videogrammetry. We treat the skeletal digits as inelastic members that provide the displacement boundary conditions for the elastic membrane. The discrete joint coordinates and Euler angles are processed through a Fourier decomposition and cubic interpolation pipeline, transforming discrete experimental snapshots into a space-time continuous variation. This allows for the generation of skeletal kinematics at the fine temporal resolution (4,000 steps per cycle) required for numerical stability in the flow solver.

\subsection{Membrane Dynamics and Spring Network Model}
We model the wing membrane as a zero-thickness shell using a spring-network approach\cite{bridson2005simulation,marco2016moving}. This method is chosen for its robustness and relative simplicity in coupling with immersed boundary solvers compared to traditional finite-element methods. The model has been successfully applied to problems ranging from bat flights (\cite{kumar2025computational} and swimming using caudal fin propulsors (\cite{kumar2026fish}. The surface is discretized into $N$ nodes and $E$ triangular elements capable of resisting both in-plane stretching and out-of-plane bending.

The dynamics of each node are governed by Newton’s second law:
\begin{equation}
    \label{eq:ode}
    m \frac{d^2 \mathbf{x}(t)}{dt^2} = \mathbf{f}(\mathbf{x})
\end{equation}
where $m$ is the nodal mass and $\mathbf{x}(t)$ is the instantaneous position. The total force $\mathbf{f}(\mathbf{x})$ is a summation of external aerodynamic loads ($\mathbf{f}_\text{ext}$), internal elastic and bending forces ($\mathbf{f}_\text{int}$), and viscoelastic damping ($\mathbf{f}_\text{damp}$). The assembled system for the entire mesh is:
\begin{equation}
    \label{bigode}
    M \frac{d^2 \mathbf{X}(t)}{dt^2} = \mathbf{F}_\text{ext} - \mathbf{F}_\text{int} - \zeta \frac{d\mathbf{X}(t)}{dt}
\end{equation}
where $M$ is the diagonal mass matrix and $\zeta$ is the structural damping coefficient, chosen to minimize numerical oscillations while allowing physical membrane response.

\subsubsection{Force Formulations}
The external force $\mathbf{F}_\text{ext}$ is derived from the fluid stresses (pressure and shear) acting on the triangular elements. For a node $i$ connected to $n_e$ elements, the force is:
\begin{equation}
    \label{fext}
    \mathbf{F}_{\text{ext},i} = \sum_{j=1}^{n_e} \frac{1}{3} \left( -\Delta p \mathbf{I} + \Delta\mathbf{\tau}\right) \cdot \mathbf{n} A^e_j
\end{equation}
where $\Delta p$ and $\Delta \tau$ represent the pressure and shear stress differences across the membrane, and $A^e$ is the element area.

Internal forces $\mathbf{F}_\text{int}$ are partitioned into elastic ($\mathbf{F}_\text{elas}$) and bending ($\mathbf{F}_\text{bend}$) components. Elastic forces follow a linear spring model based on edge deformation. Bending forces are calculated using the Helfrich energy approach\cite{bridson2005simulation,fedosov2010systematic}, which considers the angle $\theta$ between adjacent triangular elements $e_1$ and $e_2$:
\begin{equation}
    \mathbf{F}^i_\text{bend} = k_b \frac{|\mathbf{l}_e|^2}{|\mathbf{n}_1|+|\mathbf{n}_2|} \left[ \sin\frac{\theta}{2} - \sin \frac{\theta_0}{2} \right] s_i
    \label{eq:bendingForce}
\end{equation}
The stiffness coefficients $k_e$ and $k_b$ are functions of the Young’s modulus ($E^*$), membrane thickness ($h^*$), and Poisson's ratio ($\nu$).

\subsection{Fluid-Structure Interaction (FSI) Coupling}
The flow and structural solvers are coupled via an explicit (sequential) mechanism. At each time step, the flow solver computes the surface pressure and shear stresses based on the current wing configuration and passes them to the structural solver. The structural solver integrates Equation \ref{bigode} using the Newmark-beta scheme to determine the new nodal positions and velocities. These velocities are then returned to the flow solver to serve as the boundary conditions for the subsequent time step. This coupling strategy is particularly effective for the high solid-to-fluid density ratios ($\rho^* \approx 1000$) characteristic of bat flight in air.

\begin{figure}[H]
    \centering
    \includegraphics[width=1\textwidth]{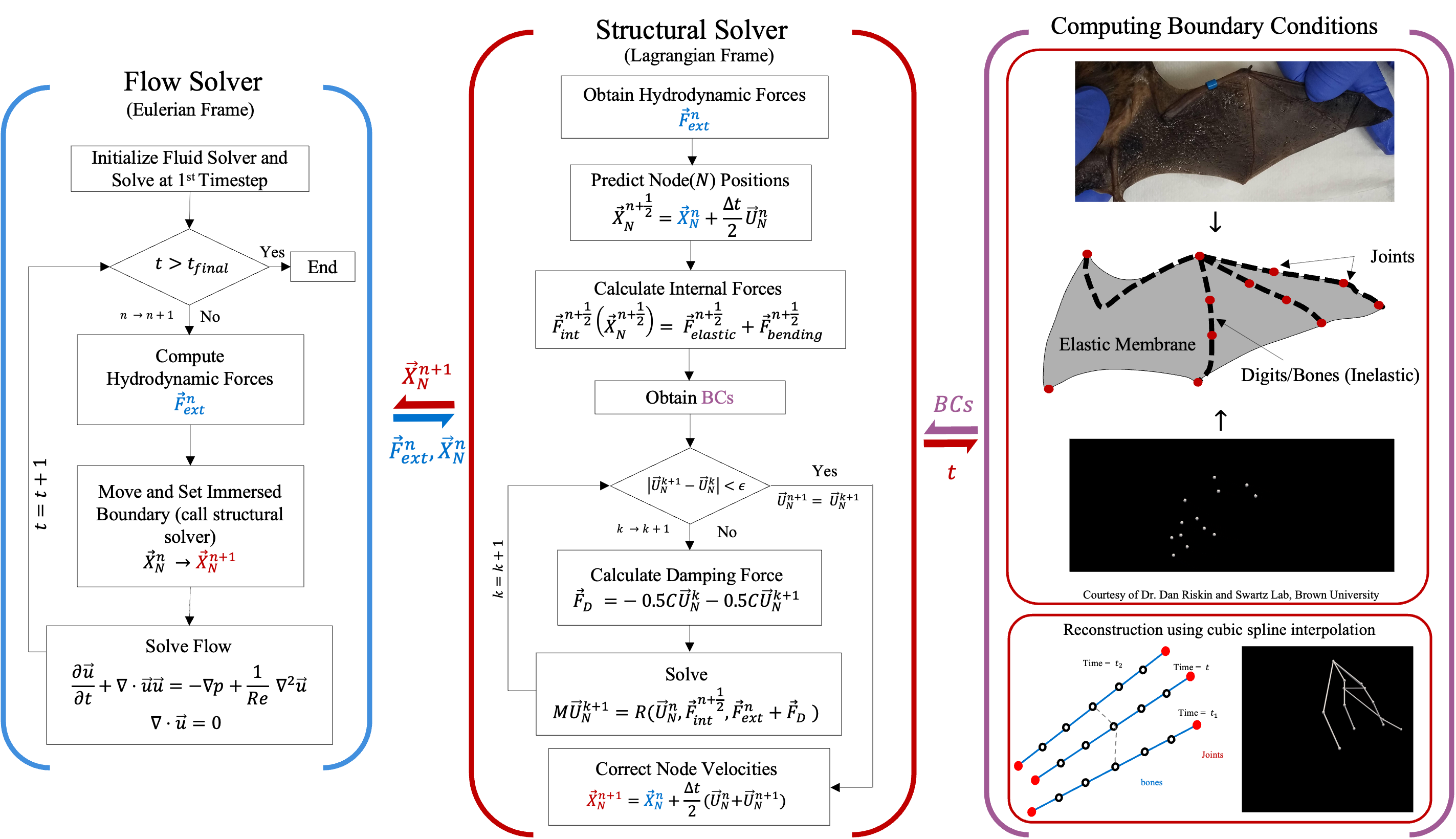}
    \caption{Flow chart showing coupling mechanism of flow and membrane dynamics solvers}
    \label{fig:flowchart}
\end{figure}

\subsection{GPU Porting}
ViCar3D consists of the following three critical modules:
\begin{enumerate}
    \item Voxel segmentation and rasterization - The subroutines within this model are responsible for classifying the grid domain represented through voxels into the interior and exterior of the body. We will refer to this raterization as ``iblanking'' in the later sections of this document.
    \item GCM modules - The subroutines handle applying the internal boundary conditions, such as no-slip and no-penetration boundary conditions. The numerical methodology is described in Mittal et al. \cite{mittal2008versatile}.
    \item Sparse Linear Solvers - These subroutines are responsible for obtaining the solution of the sparse linear solver formed from the advection diffusion equation (AD) and the pressure Poisson equation(PPE). We utilize a matrix-free and ELLPack dense matrix to efficiently handle the time-varying sparse coefficient matrix structure.
\end{enumerate}
Each routine performs specialized tasks to directly ingest a triangulated body mesh and yield a CFD solution of flow. About 140 logic loops enable this pipeline and porting and validating this pipeline is an extensive task. We utilize interoperability between OpenACC and CUDA to reprogram the entire codebase to be entirely GPU-native efficiently. The simpler nested loop structures are handled by directive-based parallelization using OpenACC, and we wrote CUDA kernels to port complex parts of the code. We also utilize NVIDIA libraries, such as Thrust and cuBLAS, to efficiently handle operations like array compaction and batch matrix inversions. 

Even though a single GPU offers CUDA cores equivalent to 100s of CPUs, the on-chip memory becomes a bottleneck in solving large systems. We therefore utilize multiple GPUs, numerically connected through halo regions, to distribute the computational problem. This not only enables us to tackle large problems but also provides additional computing power to solve them more efficiently. We employ a combination of MPI and NCCL to allow GPU-aware communication and avoid slow CPU staging of buffer data. Additionally, we utilize non-blocking and overlapped communication to hide the latency of computation when communicating over slower interconnects.

\section{Results and Discussions}
Here, we present the performance testing of our developed solver and extension to bat wing FSI problem. It is important to note that in the case of moving boundary problems involving incompressible flows, iblanking and PPE are the two most expensive modules. The AD system, due to its diagonal dominant construction, is relatively faster to solve. First, we will focus on scaling the code to multi-GPU platforms, which is achieved by hiding communication latency behind sections of the code that are independent of the data being communicated, as shown in Figure \ref{fig:profiler}. This section becomes essential when the interconnect between GPU nodes is a big bottleneck. This modification significantly improved the code's scalability. Next, we present the strong and weak scaling results performed on the GPU-resources through JHU ARCH Rockfish computer. We utilize PCIe based NVIDIA A100 with each node consisting of 4 of these. The nodes are connected with HDR100 interconnect which theoretically split the 100Gbps or 12.5 GBps between the 4 GPUs.

\subsection{Performance Testing}
\begin{figure}[hbt!]
  \centering
  \begin{subfigure}[b]{0.47\textwidth}
    \centering
    \includegraphics[width=\textwidth]{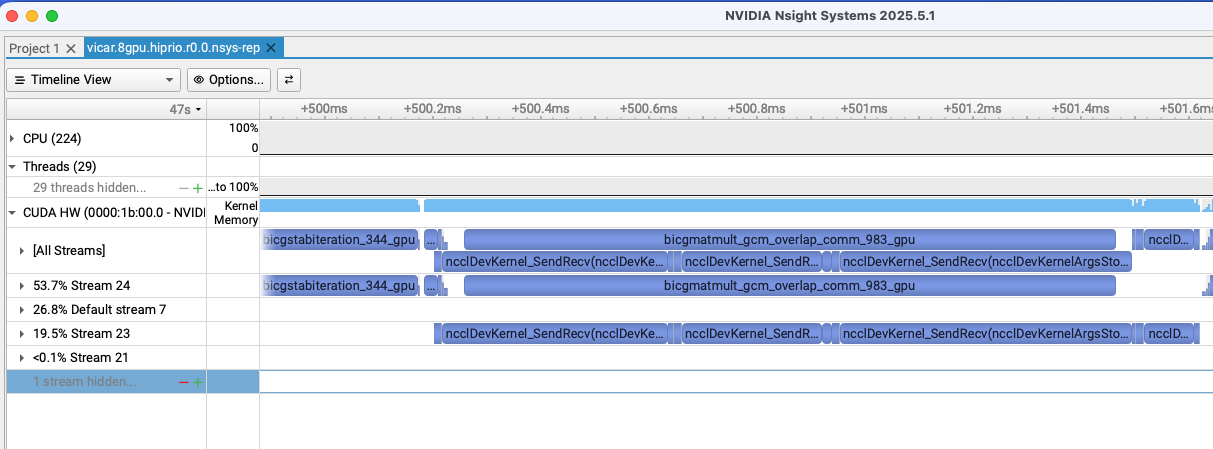}
    \caption{}
    \label{fig:profiler}
  \end{subfigure}
  \quad
  \begin{subfigure}[b]{0.49\textwidth}
    \centering
    \includegraphics[width=\textwidth]{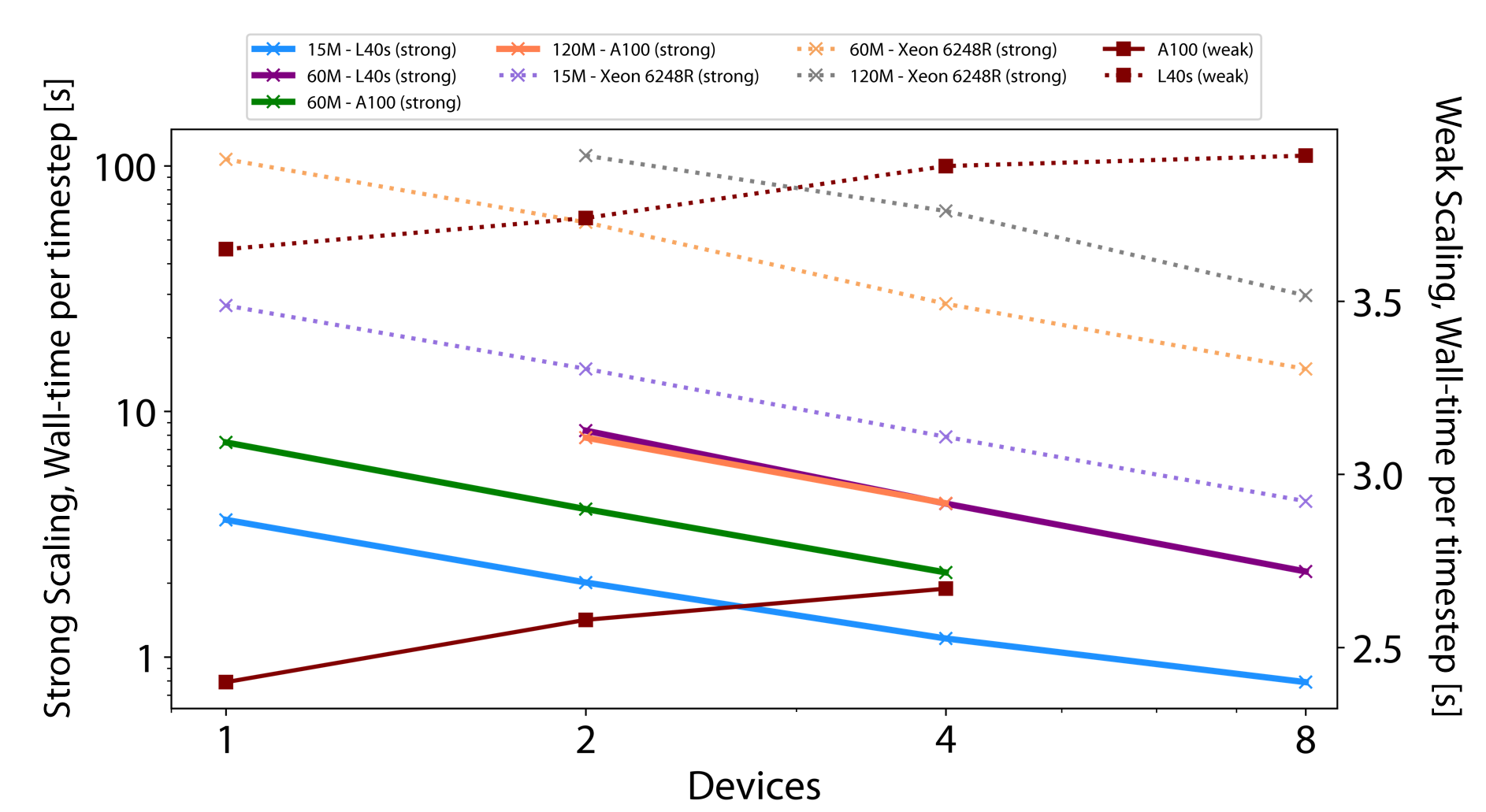}
    \caption{}
    \label{fig:scaling}
  \end{subfigure}
  \caption{Scaling to multiple GPUs (a) Overlapped Communication to Computation visualized through \texttt{Nsight System Profiler}, (b) Scaling plot showing both strong and weak scaling}
  \label{fig:domain}
\end{figure}

We utilize weak and strong scaling as two measurements for our performance testing which are presented in figure \ref{fig:scaling}. Weak scaling assesses the solver's ability to maintain performance as the problem domain expands in proportion to the number of GPUs. We fixed the per-GPU subdomain resolution at $301 \times 385 \times 129$ grid points—approximately 15 million points per GPU in double precision for simulating the flow around a 3D rectangular wing at a Reynolds number of Re = 1000. The setup is adapted from Menon et al. \cite{menon2022contribution} but with pitching wing undergoing motion at $St = 0.1$ nad $\alpha = 5^\circ$. The global mesh was then scaled linearly up to 4 GPUs on the A100 platform and 8 GPUs on the L40S platform. On the A100 (40 GB memory per GPU), the per-time-step wall time increased from 2.40 s for one GPU to 2.58 s for two GPUs and 3.67 s for four GPUs. This corresponds to a weak scaling efficiency of roughly 90\%. For the L40S platform, wall times started at 3.65 s and increased to 3.92 s across eight GPUs, achieving about 93\% efficiency. The code is thus able to sustain a weak scaling $>93\%$.

Strong scaling evaluates speedup for fixed problem sizes by adding GPUs, revealing limits in latency, throughput, and communication. We tested three baseline grids: 15M points ($301\times385\times129$), 60M points ($603\times771\times129$), and 120M points ($1205\times771\times129$). This range also enables us to probe kernel launch overheads, which become more pronounced relative to computation as the grid size decreases. For the 15M grid, as suspected, efficiency dropped to 60\% at higher GPU counts (e.g., 200,000 points per GPU on eight GPUs), highlighting communication and launch latency bottlenecks. With the 60M grid, we achieved 86\% efficiency on A100 and 90\% on L40S across up to eight GPUs. The 120M grid exceeded the single-GPU memory limit; therefore, results are presented for only 2 and 4 GPUs, yielding a 92\% normalized efficiency. This indicates that when the workload is sufficient enough to saturate the GPU resources, we observe a strong scaling efficiency $>90\%$ as well.

\begin{figure}[hbt!]
  \centering
  \begin{subfigure}[b]{0.47\textwidth}
    \centering
    \includegraphics[width=\textwidth]{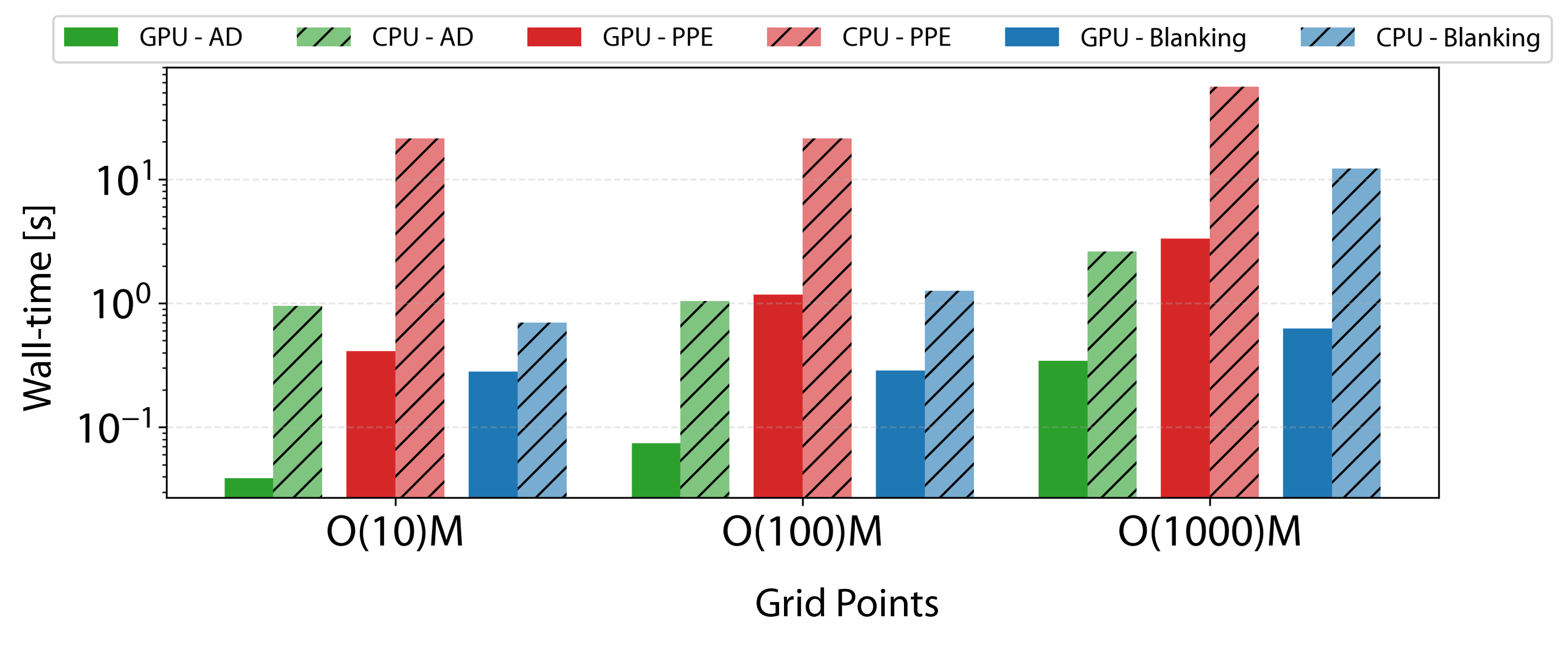}
    \caption{}
    \label{fig:element_walltime}
  \end{subfigure}
  \quad
  \begin{subfigure}[b]{0.49\textwidth}
    \centering
    \includegraphics[width=\textwidth]{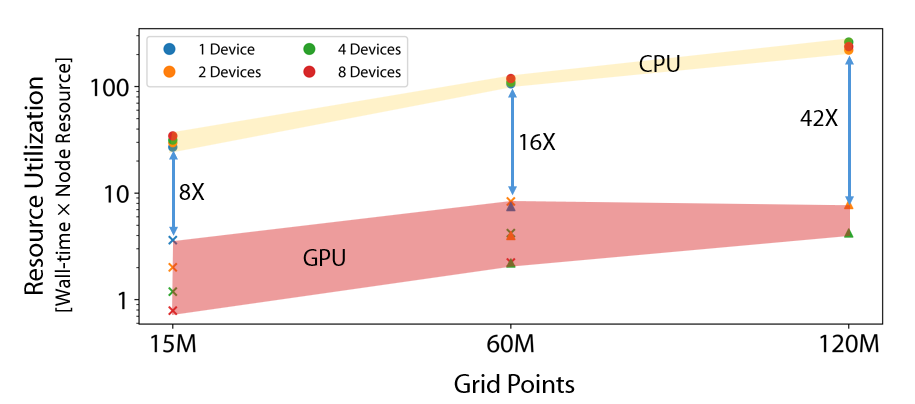}
    \caption{}
    \label{fig:resource_utilization}
  \end{subfigure}
  \caption{(a) Speed up of dominant subroutines within the code (b) Resource savings with GPU utilization}
  \label{fig:domain}
\end{figure}

Next, we focus on the elemental speedup for the compute dominant elements within the code as presented in figure \ref{fig:element_walltime} for three order of grid points : $O(10 \,\text{million})$, $O(100 \,\text{million})$ and $O(1000 \, \text{million})$/$O(1 \, \text{billion})$ . We observe a consistent speed of an order-of-magnitude across the range of grid points. In figure \ref{fig:resource_utilization} we present the resource savings by utilizing the GPU hardware. The resource utilization is computed by multiplying the the number of devices with the wall-time to perform a specific calculation timestep. This unit of measurement is the actual charge user has to pay for usage on HPC cluster. We report results for the metric from simulations ranging from 15M to 120M. The test was done for resource upto 8 silicon dies - 8 NVIDIA L40s GPUs and 8 Intel Xeon CPUs. It can be seen that at 15M grid simulation we are obtaining a 8X resource saving, which increases to 42X when going to a 120M grid simulation.

These results shows that the GPU adaptation of the immersed boundary method have enabled massive speedups along with resource saving for moving boundary problems. Readers are encouraged to read our previous work Kumar et al. \cite{kumar2026gpu} for verification test results.

\subsection{Fluid-Structure Interaction in Bat}

In this section, we describe a simulation of fully coupled aero-structural dynamics of a bat-inspired membrane wing in forward flight. Simulating this bio-inspired locomotion is highly complex due to the massive geometric deformations caused by the interplay of pitching, flapping, active skeletal articulation, and flow-induced membrane deformation. The wing model consists of an elastic membrane supported by a highly articulated skeleton, comprising four primary segments: the propatagium, plagiopatagium, dactylopatagium major, and dactylopatagium medius. Readers are encouraged to read about the computational framwework to perform these realistic bat flight simulation from our earlier work in Kumar et al. \cite{kumar2025computational}. We keep the simulation parameters especially the Advance ratio same to have similarity with the experiments. The only difference from our prior work is the increase in Reynold number from 1000 to 5000 which is expected to generate multiscale and complex vortex structures. 

To capture the complex vortex shedding and boundary layer interactions at a Reynolds number of $Re = 5,000$, we employ a massive Cartesian grid of $750 \times 754 \times 800$, corresponding to approximately 452 million grid points. A uniform, highly resolved grid spacing of $0.0035$ is maintained in a cuboidal region around the body, which stretches rapidly toward the outer boundaries to dampen far-field acoustics. This minimum grid resolution places about 300 grid points necessary to accurately capture vortex structures for correct quantification of hydrodynamic forces. A symmetry boundary condition is applied at the wing root to mimic a dual-wing configuration and prevent unphysical root vortices. The simulation timestep was set to $\Delta t = 0.0005$ to maintain a stable CFL number during the severe membrane accelerations of the flapping cycle. This extreme-scale FSI simulation was performed on 1 nodes utilizing 4 NVIDIA RTX Blackwell Max-Q GPUs. The simulation required approximately 60 hours to complete 1 flapping cycles (up to $t/T = 4.0$). 

\begin{figure}[H]
\centering
\includegraphics[width=0.9\textwidth]{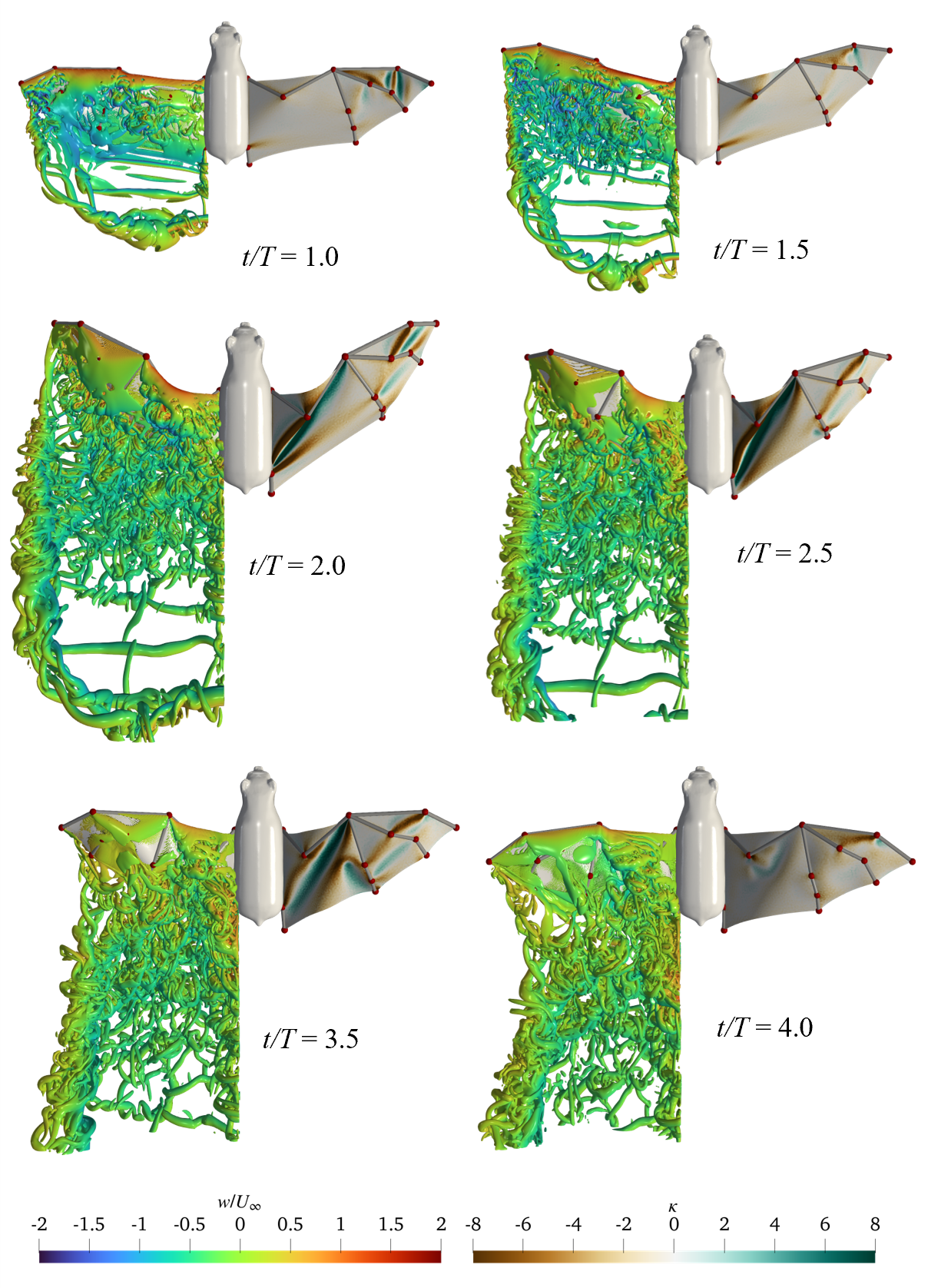}
\caption{Evolution of the complex wake and aero-structural dynamics over four flapping cycles ($t/T = 1.0$ to $4.0$). The left wing displays $Q$-criterion isosurfaces coloured by vertical velocity ($w/U_\infty$), highlighting the formation of the LEV, tip vortices, and the transition into a dense wake. The right wing shows the corresponding structural configuration, highlighting the localized deformation using curvature as a metric.}
\label{fig:isosurfaces}
\end{figure}

Having established the computational setup, we now focus on the vortical features generated as the highly articulated wing interacts with the flow during the flapping cycle. Figure \ref{fig:isosurfaces} shows snapshots of the vortex structures generated at key phases from $t/T = 1.0$ to $4.0$, visualized using isosurfaces of $Q$-criterion coloured by the vertical velocity. It is to be noted that we had only simulated left wing. The body and the right wing showing wing deformation are added for improved visualization. It can be seen that the  GPU-accelerated computing enables to simulate such a large case with minimal GPU resources. It essentially enabled us to bring HPC scale CPU simulation to local consumer GPU server. Moreover, it allows for the detailed capture of these complex vortical features and their breakdown over the flapping cycle. In this work, we are focussed on the first flapping cycle as the goal is capability demonstration and not detailed insight into the flow physics.

The downstroke starts with the formation of leading-edge vortices along the leading edges of the propatagium and dactylopatagium medius due to the interaction of incoming airflow and the downward-moving leading edge of the membrane wing. The LEV over the propatagium sheds periodically during the downstroke, forming a series of turbulent wake vortices. Concurrently, the tip vortex (TV) that formed during the previous cycle is seen detaching from the wingtip. The vortices originating from the propatagium detach from the wing and merge with the trailing-edge vortices, giving rise to complex wake vortices that become increasingly intricate by $t/T = 4.0$.

While the LEV at the propatagium continuously forms and breaks, the LEV at the dactylopatagium major and medius remains attached. This is due to the reduced local effective angle of attack caused by the supination of this part of the wing. This LEV transitions to a triangular-shaped vortex and spans the dactylopatagium medius and major, remaining attached over the entire downstroke and the first quarter of the upstroke. The downward wing motion also results in the formation of a strong TV that is identifiable throughout the entire flapping sequence.

\begin{figure}[H]
  \centering
  \begin{subfigure}[b]{0.48\textwidth}
    \centering
    \includegraphics[width=\textwidth]{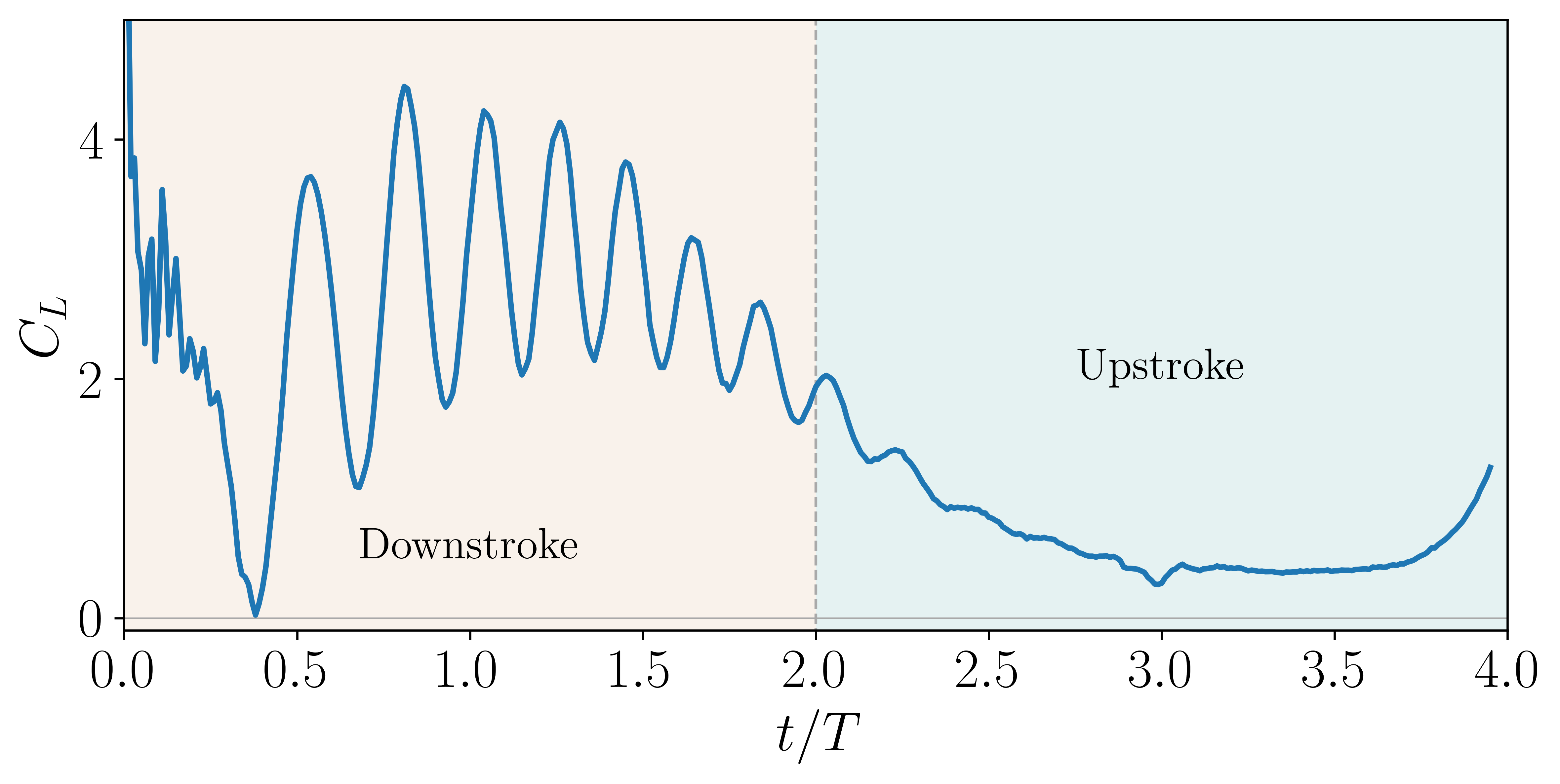} 
    \caption{Lift Coefficient ($C_L$)}
    \label{fig:cl_history}
  \end{subfigure}
  \hfill
  \begin{subfigure}[b]{0.48\textwidth}
    \centering
    \includegraphics[width=\textwidth]{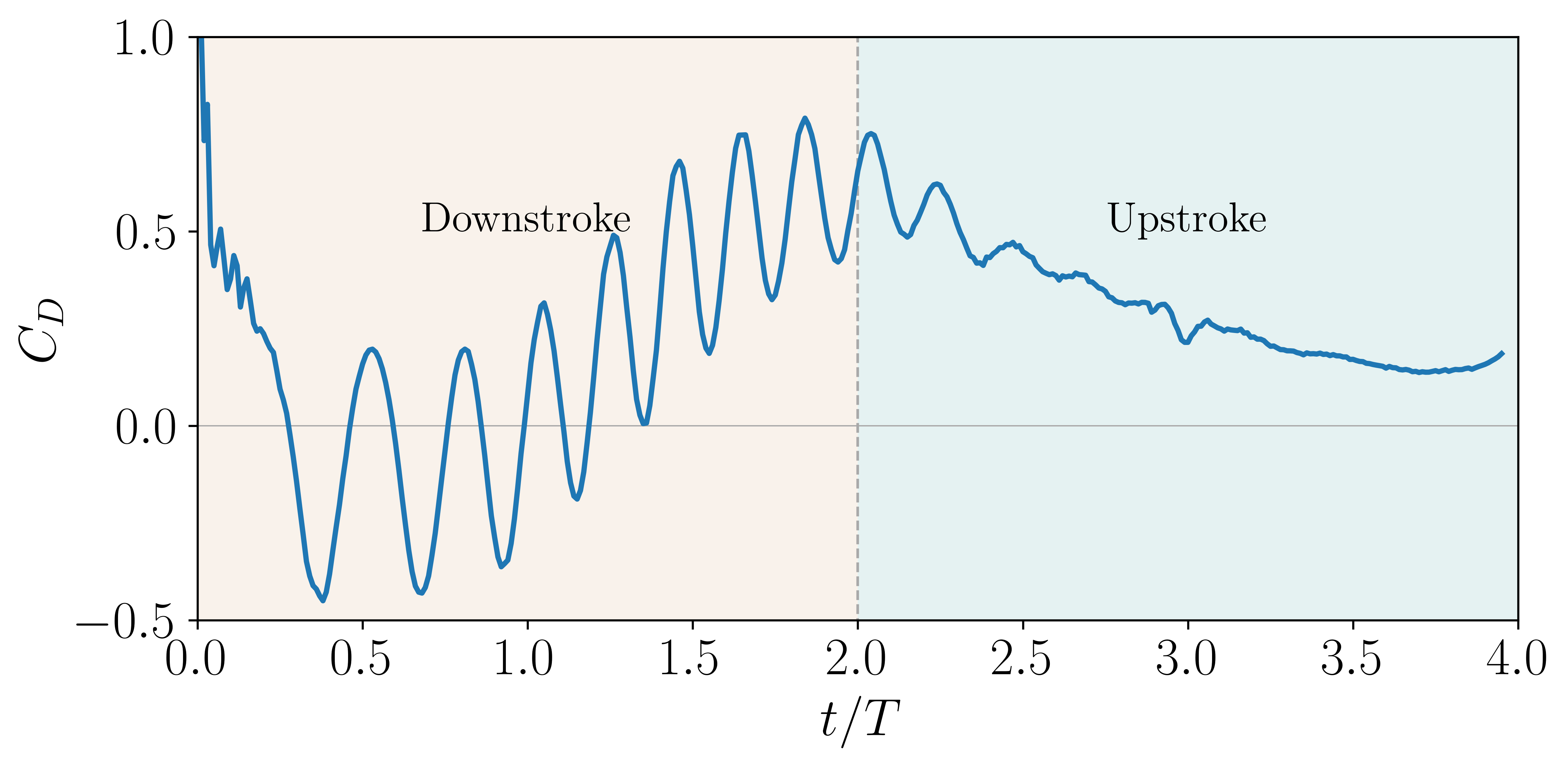} 
    \caption{Drag Coefficient ($C_D$)}
    \label{fig:cd_history}
  \end{subfigure}
  \caption{Time variation of the aerodynamic forces over four flapping cycles. The plots highlight the highly oscillatory, power-generating downstroke compared to the aerodynamically smooth recovery upstroke.}
  \label{fig:aeroforces}
\end{figure}

We focus here on the generation of lift and drag forces and quantify them in terms of the time histories of the overall lift ($C_L$) and drag ($C_D$) coefficients. As illustrated in Figure \ref{fig:aeroforces}, the flapping cycle initiates with the bat performing the lift-generating downstroke, followed by the upstroke or the recovery stroke. 

During the downstroke, the wing generates the vast majority of the aerodynamic forces, with major contributions to the lift generated during this phase. The time traces exhibit severe, high-frequency load fluctuations. These oscillations are a direct manifestation of the fully coupled fluid-structure interaction, reflecting the dynamic shedding of leading-edge vortices and the highly unsteady aeroelastic vibrations of the flexible wing membrane. The plagiopatagium segment of the wing is most effective in generating drag, mostly at the end of the downstroke where the wing has a large angle of attack and the flow is separated. The propatagium and dactylopatagium medius, especially its leading-edge region, is singularly effective in generating thrust, and this is also due to the formation of the LEV during the downstroke when the ventral surface of this segment pitches down.

As the wing transitions into the upstroke, the plagiopatagium experiences large bending strain, which peaks as the phalanges are retracted resulting in the development of slack in the membranes. Throughout the upstroke, the aerodynamic force profiles become notably smoother. Furthermore, the articulated wing is able to sustain positive lift during portions of the upstroke, which is connected with the articulation of the wing.

\section{Conclusion}
In this work, we have developed and assessed a highly optimized, GPU-native sharp-interface immersed boundary method capable of simulating extreme-scale fluid-structure interaction problems. By leveraging a hybrid parallelization strategy that combines OpenACC, custom CUDA kernels, and NCCL-based communication, we successfully mitigated the traditional memory and latency bottlenecks associated with high-resolution sharp interface Cartesian grids-based methods. Performance profiling demonstrates a $20\times$ wall-clock speedup and up to a $42\times$ reduction in computational resource costs compared to equivalent multi-core CPU architectures, all while maintaining $>90\%$ strong and weak scaling efficiencies across multiple modern GPU platforms. 

We demonstrated the robustness and high-fidelity capabilities of this solver by performing a fully coupled aero-structural simulation of a bat-inspired membrane wing in forward flight at a Reynolds number of 5,000 with a massive grid of approximately 452 million points. The simulation was able to capture the complex multi-scale vortex dynamics such as the dynamic shedding of leading-edge vortices and the resulting high-frequency aeroelastic load fluctuations over the flapping cycle. Crucially, the computational efficiency of our GPU framework allowed this massive FSI problem to be solved on a localized consumer-grade GPU server with 4 RTX Blackwell Pro Max-Q GPUs in just 60 hours per flapping cycle. Ultimately, this work brings historically prohibitive extreme scale simulations of biological flight simulations into a practical computational time-frame. Moreover, the solver is highly versatile and is being used to study FSI in physiological configurations, marine biolocomotion, and in aeroelastic and hydroelastic systems.

\section*{Acknowledgments}
The development of the GPU solver benefited from NSF grant - CBET-2011619, ONR grant - N00014-22-1-2655 and NVIDIA academic research grant. The work benefited from the computational resources provided by Advanced Research Computing at Hopkins Rockfish supercomputer and DoD HPC Modernization Program's AFRL Raider supercomputer.

\bibliography{sample}

\end{document}